\documentstyle[epsfig]{mn}
\begin{document}


\title[Cygnus X-3 in outburst]
{Cygnus X-3 in outburst :
quenched radio emission, radiation losses and variable local opacity}

\author[R. P. Fender et al.]
{R.P.~Fender,$^{1,2}$
S.J.~Bell~Burnell,$^1$ 
E.B.~Waltman,$^3$ 
G.G.~Pooley,$^4$
F.D.~Ghigo$^5$ \cr
and R.S.~Foster$^3$\\
$^1$ Department of Physics, The Open University, Walton Hall, Milton Keynes
MK7 6AA \\
$^2$ Astronomy Centre, University of Sussex, Falmer, Brighton BN1 9QH\\
$^3$ Remote Sensing Division, Naval Research Laboratory, Code 7210,
Washington, DC 20375-5351\\
$^4$ Mullard Radio Astronomy Observatory, Cavendish Laboratory, 
Madingley Road, Cambridge CB3 0HE\\
$^5$ National Radio Astronomy Observatory, P.O. Box 2, Green Bank, West
Virginia 23944\\
}

\maketitle

\begin{abstract}

We present multiwavelength observations of Cygnus X-3 during an
extended outburst in 1994 February - March. Intensive radio monitoring
at 13.3, 3.6 \& 2.0 cm is complemented by observations at
(sub)millimetre and infrared wavelengths, which find Cyg X-3 to be
unusually bright and variable, and include the first reported
detection of the source at 0.45 mm. We report the first confirmation
of quenched radio emission prior to radio flaring independent of
observations at Green Bank. The observations reveal evidence for
wavelength-dependent radiation losses and gradually decreasing opacity
in the environment of the radio jet.  We find that the radiation
losses are likely to be predominantly inverse Compton losses
experienced by the radio-emitting electrons in the strong radiation
field of a luminous companion to the compact object. We interpret the
decreasing opacity during the flare sequence as resulting from a
decreasing proportion of thermal electrons entrained in the jet,
reflecting a decreasing density in the region of jet formation.  We
present, drawing in part on the work of other authors, a model based
upon mass-transfer rate instability predicting $\gamma$-ray, X-ray,
infrared and radio trends during a radio flaring sequence.

\end{abstract}


\section{Introduction}

Cygnus X-3 was discovered as a strong X-ray source by a rocket flight
in 1966 (Giacconi et al. 1967).  Six years later in 1972 it was found
to have a radio counterpart (Braes \& Miley 1972).  In that same year
Cyg X-3 underwent its first observed series of radio flares,
attracting an intense radio monitoring campaign (see Gregory et al.
1972 et seq). During this period X-ray observations detected a 4.8
hour modulation in the signal from Cyg X-3 (Parsignault et al. 1972;
Sanford \& Hawkins 1972); this was interpreted as being the orbital
period of the system. A 12th mag infrared K-band counterpart was
discovered using the radio position (Becklin et al. 1972), and this was
subsequently found to exhibit the same 4.8 hour modulation observed in
X-rays (Becklin et al. 1973), confirming the association between radio,
IR and X-ray sources.

In more recent years, as well as regular monitoring at cm wavelengths
(e.g. Waltman et al. 1994, 1995), Cyg X-3 has been observed at mm
wavelengths (Baars et al. 1986; Fender et al. 1995), in the infrared
between 0.9 - 4.8 $\mu$m (e.g. Molnar 1988, van Kerkwijk et al. 1996,
Fender et al. 1996), and in the red-optical (Wagner et al. 1990).  In
the visible band Cyg X-3 remains undetected due to the high degree of
interstellar absorption along the line of sight to the system ($A_V >
20$ mag : e.g.  van Kerkwijk 1996) At higher energies Cyg X-3 is a
bright X-ray source with luminosity states which seem to be correlated
with radio activity (Watanabe et al. 1994).  In the high state the
X-ray luminosity in Cyg X-3 approaches, and possibly exceeds, the
Eddington limit for a solar mass object. Cyg X-3 is also a strong
source of $\gamma$-rays, with claimed detections at up to PeV energies
(see e.g. Bonnet-Bidaud \& Chardin 1988; Protheroe 1994 for reviews).
Radio mapping observations have revealed the presence of a
milliarcsecond-scale bipolar jet with an expansion velocity of $\sim
0.35$ c (at 10 kpc) following both giant- and small-flaring periods
(Geldzahler et al. 1983; Spencer et al. 1986; Molnar, Reid \& Grindlay
1988; Schalinkski et al. 1995) as well as more extended structures on
arcsecond- (Strom, van Paradijs \& van der Klis 1989) and
arcminute-scales (Wendker, Higgs \& Landecker 1991).

Since 1972, of the order of 40 radio flare events have been observed
in Cyg X-3, with peak flux densities ranging from 1 to 20 Jy, in
comparison with a quiescent level of 50-100 mJy at cm wavelengths
(e.g. Waltman et al. 1994, 1995) The Autumn 1972 flare remains the
event most extensively covered at radio wavelengths (Gregory et al.
1972 et seq) and displays many properties typical of radio flaring
periods in Cyg X-3. These include a tendency for shorter wavelength
emission to peak earlier and higher than at longer wavelengths, and
peak emission initially at $\sim 6$ cm but progressing gradually
toward longer wavelengths throughout the series of flare events. The
former point is as predicted for an expanding synchrotron-emitting
cloud of relativistic electrons by van der Laan (1966).  The latter
point is harder to explain, and although it may be considered to be
due to a general decrease of opacity at cm wavelengths in the local
environment of the emitting electrons (Waltman et al. 1995) exactly
what this interpretation really means is unclear.  General consensus,
however, for models of the radio flaring emission does involve a
synchrotron-emitting cloud of electrons, probably associated with the
radio jets, and with varying degrees (and causes) of local opacity and
geometry (e.g. Gregory et al. 1972; Marscher \& Brown 1975; Mart\'{\i},
Paredes \& Estalella 1992 - hereinafter MPE92).  Observations in bands
other than the radio during periods of radio flaring have been rare,
though Pomphrey \& Epstein (1972) and Baars et al. (1986) have observed
mm emission during active radio periods and found it to be roughly
consistent with the synchrotron tail of the cm emission (although
Baars et al. found the rise and decay times of flares at 3.3 \& 1.3 mm
to be much shorter than their cm counterparts); and the strong-lined
infrared spectrum reported in van Kerkwijk et al. (1992) was obtained
during the decay period of a 4 Jy flare (Kitamoto et al. 1994).

\begin{figure*}
\begin{minipage}{177mm}
\centering
\leavevmode\epsfig{file=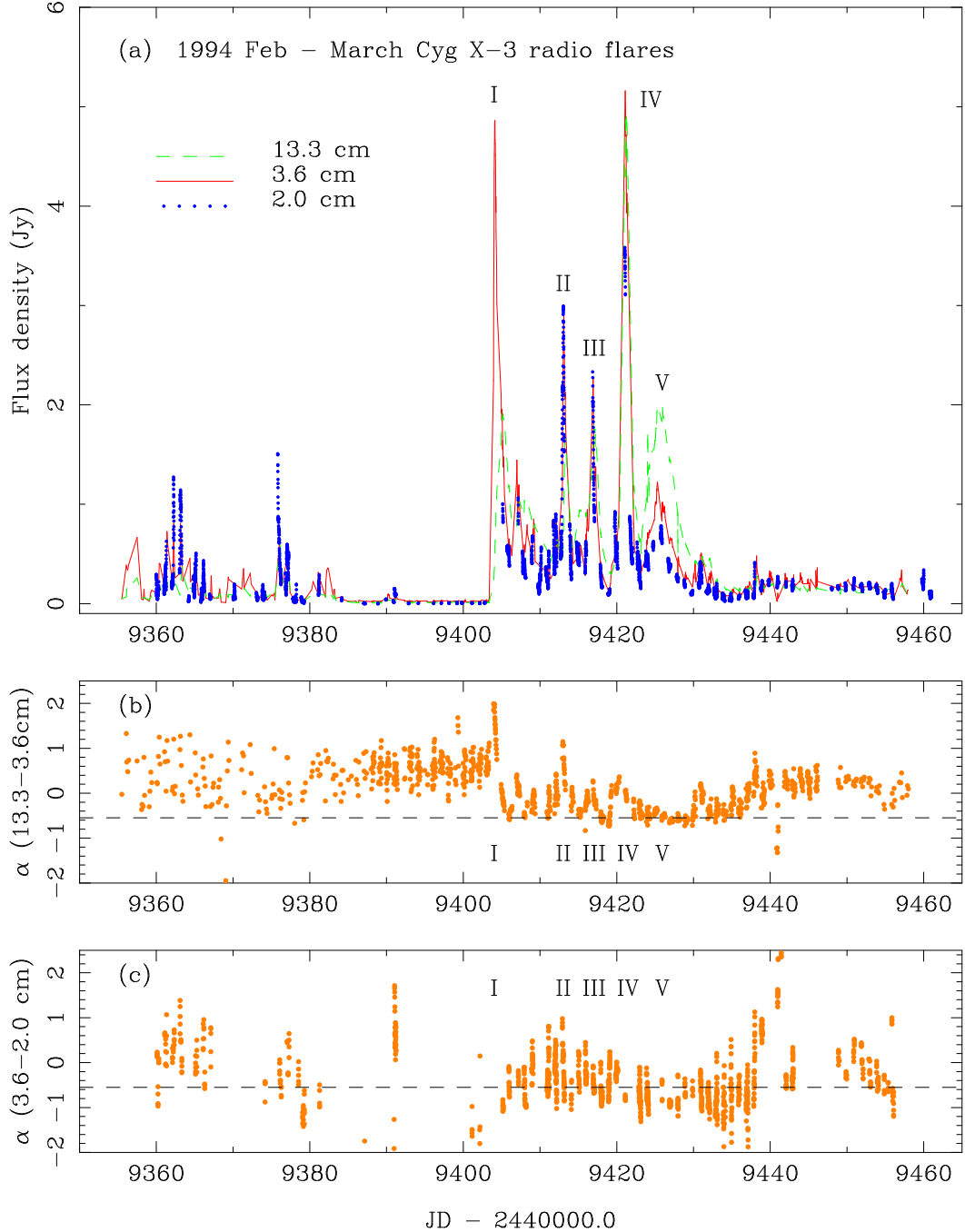, width=14cm, clip}
\caption{Monitoring of the 1994 Feb - March radio flares at 13.3, 3.6 \&
2.0 cm. The five major flare events have been labelled I - V. Figs 1(b)
\& 1(c) show how the spectral indices from 13.3 - 3.6 cm \& 3.6 - 2.0 cm
respectively evolve with time}
\label{}
\end{minipage}
\end{figure*}

\section{Observations}

During the 1994 Feb - Mar flaring period Cyg X-3 was monitored extensively
at 2.0 cm with the Ryle Telescope at Cambridge, and at 3.6 and 13.3 cm
with the NRL-Green Bank Interferometer. These observations are illustrated
in Fig 1; we consider there to have been five flare events during this
period and we label these I - V. Observations were also obtained
in the infrared using UKIRT, and at mm and sub-mm wavelengths using JCMT,
both on Mauna Kea in Hawaii.

\subsection{Ryle Telescope}

The Ryle Telescope (Jones 1991) has been used to monitor variable
sources, primarily during gaps in the main observing programmes of the
telescope, in addition to the coordinated observations which were
specifically scheduled.

During the period discussed here, the telescope operated at 15 GHz
with a bandwidth of 350 MHz.  The Stokes' parameters I+Q were measured
in all cases.  Data from each interferometer pair are collected,
together with interleaved calibration observations of B2005+403, a
nearby quasar which is unresolved at all baselines in use here.  The
calibrations were used to correct for the instrumental phases of the
system; the amplitude scale was calibrated by nearby observations of
3C48 or 3C286. Atmospheric refractive-index fluctuations introduce
phase errors which increase with baseline; the time-scales of these
are such that it is not possible to remove them by using the
interleaved calibrator. The optimum procedure is to use only the
shorter baselines, for which the phase uncertainties are small, except
in the few cases where the flux is exceptionally low (before major
flares) when it may be desirable to use all the available data. During
the observations described here, the telescope was in its `compact
array' mode with 5 aerials in a group having maximum baseline of about
100m; in practice nearly all data use only this set of baselines.

Once the instrumental phases have been removed, the data for all the
baselines selected are added, producing the effect of a phased
array. This vector sum is smoothed in time and its amplitude
is taken as the flux of the source (in the rare cases of very low flux
density, this procedure overestimates the flux; the in-phase component
of the vector sum is used to give an unbiased estimate). Typical 
day-to-day uncertainty in the flux-density scale is less than 3\% rms.
The noise level on a typical 5-minute average is less than 2 mJy rms.

\subsection{NRL-Green Bank Interferometer}

Cyg X-3 has been observed by the NRL-GBI Monitoring Program on a daily
basis since 1982. The Green Bank Interferomter (GBI) was operated by
the National Radio Astronomy Observatory for the Naval Research
Laboratory (NRL) and the US Naval Observatory through 1996 April 1.
In Autumn 1989, the National Radio Astronomy Observatory installed
cryogenic receivers on the Green Bank Interferometer.  Since October
1989, Cyg X-3 and the calibration sources were observed for
approximately ten minutes each, simultaneously at two frequencies,
2.25 and 8.3 GHz.  The flux densities were measured on the 2.4 km
baseline and are the average of the left-left and right-right circular
polarizations.  Cyg X-3 was observed three to 15 times a day on two
different gain codes to facilitate accurate measurements over a range
of possible, and occasionally rapidly changing, flux densities from
100 mJy to almost 20 Jy.

A flux density calibration procedure similar to that reported in
Waltman et al. (1994,1995) was employed here.  Three secondary
calibration sources (B0237$-$233, B1245$-$197, and B1328+254) were
used to produce flux densities for Cyg X-3 and B2005+403.  The flux
densities of B0237$-$233, B1245$-$197, and B1328+254 were determined
using observations of B1328+307 (3C 286).  The flux density of 3C 286
was based on the scale of Baars et al. (1977), and the assumed values
were 11.85 Jy at 2.25 GHz and 5.27 Jy at 8.3 GHz.

The quiescent flux densities for Cyg X-3 were calibrated using the
four calibration sources listed above, weighted by the difference in
time between the observations of Cyg X-3 and calibrator sources
observed within 24 hours of the Cyg X-3 observation.  However, during
flares when Cyg X-3 was observed almost continuously from 5 hours east
to 5 hours west of the meridian, scans of Cyg X-3 were paired with
scans of 2005+403 in order to remove hour angle gain effects in the
flux densities.  These may exceed 20 per cent at 8.3 GHz.  In this
case, the meridian observation of 2005+403 was calibrated using the
three calibrators above.  Then, paired scans with 2005+403 were used
to calibrate Cyg X-3.

Errors in the GBI data are flux dependent: 4 mJy (2 GHz) or 6 mJy (8
GHz) for fluxes $<$ 100 mJy, 15 mJy (2 GHz) or 50 mJy (8 GHz) for
fluxes $\sim$ 1 Jy, and 75 mJy (2 GHz) or 250 mJy (8 GHz) for fluxes
$\sim$ 5 Jy (one sigma).

\subsection{UKIRT}

Following the observations of radio flaring activity in Cygnus X-3 as
observed in the Green Bank and Ryle Telescope data,
target-of-opportunity observations were requested and made on two
nights at the United Kingdom Infrared Telescope (UKIRT). All
observations were made using IRCAM2, a 58 $\times$ 62 imaging array
operating in the $1 - 5 \mu$m wavelength range. On 1994 February 23,
less than five days after the onset of the first radio flare,
observations were made twice each in the H (1.6 $\mu$m) and K (2.2
$\mu$m) bands with an integration time of 10 s and flux calibrated
using HD136754. On the following night a series of five observations
in the K band were made, each with 15 s integration time and flux
calibrated using HD162208. The observations on both nights were made
near the end of the shift as dawn approached, hence the larger than
usual errors. The UKIRT observations are summarised in table 1; there
is approximately 25 sec between subsequent exposures on the same night.

\subsection{JCMT}

We obtained, over a three day period between the first and second
major flares in the sequence, a series of (sub)millimetre observations
using the James Clerk Maxwell Telescope (JCMT). All observations were
made using UKT14, a photometer operating in the 0.45 - 2.0 mm
wavelength range. On 1994 Feb 26 we obtained a 3$\sigma$ upper limit
at 1.1 mm and shortly afterwards a 3.9$\sigma$ detection of Cyg X-3 at
0.8 mm.  Observations on the following two days at 0.45 mm provided
initially a 3$\sigma$ upper limit and then a 4.8$\sigma$
detection. The detection at 0.45 mm is the first reported detections
of Cyg X-3 at this wavelength; detection at 0.8 mm (at a similar flux
level) has also only been reported once before (Tsutsumi et al.. 1996).
The JCMT observations are summarised in table 2.

\begin{table}
\caption{UKIRT observations during 1994 Feb -- March}
\label{symbols}
\begin{tabular}{@{}ccc}
JD - 2440000.0 & Filter & Flux density (mJy) \\
9406.76 & H & 16.1 $\pm 1.0$ \\
 --     & K & 32.5 $\pm 2.0$ \\
 --     & H & 16.8 $\pm 1.0$ \\
 --     & K & 39.5 $\pm 2.0$ \\
9407.75 & K & 25.1 $\pm 1.0$ \\
  --    & K & 23.7 $\pm 1.0$ \\
  --    & K & 23.4 $\pm 1.0$ \\
  --    & K & 21.8 $\pm 1.0$ \\
  --    & K & 19.8 $\pm 1.0$ \\

\end{tabular}
\medskip
\end{table}

\begin{table}
\caption{JCMT observations during 1994 Feb -- March}
\label{symbols}
\begin{tabular}{@{}ccc}
JD - 2440000.0 & Filter & Flux density (mJy) \\
9410.32 & 1.1 mm & $<$ 40 \\
9410.33 & 0.8 mm& 77 $\pm 20$ \\
9411.34 & 0.45 mm& $<$ 430 \\
9412.36 & 0.45 mm& 330 $\pm 69$ \\

\end{tabular}
\medskip
\end{table}

\begin{figure}
\centering
\leavevmode\epsfig{file=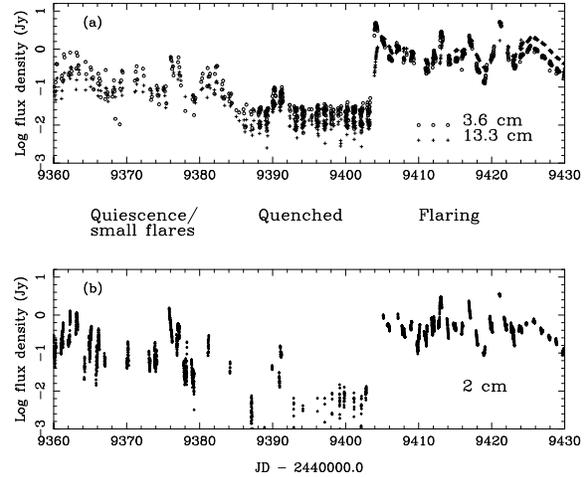, width=8.5cm, clip}
\caption{Quenched radio emission preceding flares : Fig 2(a) shows the
Green Bank monitoring revealing unusually low flux densities for $\sim 19$
days preceding flare I. Fig 2(b) presents an independent confirmation of 
this effect, in the 2.0 cm Ryle Telescope data.}
\label{}
\end{figure}

\section{The radio flares}

Here we discuss the most striking features of the observations :
confirmation of a period of very low or `quenched' flux densities
prior to radio flaring, evidence for an evolving flare spectrum and
for radiation losses during the decay phase of flares, and
observations of anomalously high and variable states in the (sub)mm
and infrared regimes.

\subsection{Confirmation of quenched radio fluxes prior to flaring}

Waltman et al. (1994, 1995) showed that radio flaring events in Cyg X-3
appear to be often, if not always, preceded by a period of quenched
emission when the flux at cm wavelengths drops to $\leq 30$ mJy
(c.f. normal `quiescent' fluxes of $50 - 150$ mJy). The 1994 Feb -
March radio flares were preceded by a long period of such quenched
radio fluxes, at least 19 days, and for the first time an instrument
other than the Green Bank interferometer was able to confirm the
effect. Observations with the Ryle Telescope during this period
revealed some of the lowest fluxes ever recorded from Cyg X-3 at radio
wavelengths, in some instances marginal detection at about 1 mJy.  Fig
2 illustrates the Green Bank and Ryle Telescope monitoring during the
period of quenching, showing the transition between quiescent/small
flaring to quenched to large flaring over a period of $\sim 30$ days.

It is interesting to note that the mean spectrum of the quenched
emission is not, however, simply optically thin from 13.3 - 2.0 cm as
would be expected if the emission arose in `old' radio lobes (Strom et
al 1989) during a period when the jet was totally supressed. Instead
there remains evidence for some ongoing opacity, particularly in the
Green Bank observations, suggesting that the quenched emission is
still arising from a reasonably dense environment local to the centre
of the binary system. We caution however that the Green Bank data are
very noisy at this level and observations during a period of quenched
emission with a more sensitive instrument (e.g. VLA) are needed to
confirm this effect.

\subsection{Evolution of flare spectra}

There are several clear trends evident in the 13.3, 3.6 \& 2.0 cm
radio data during the sequence of events I -- V. The first and most
obvious is the turn around of the radio spectrum : i.e. flare I peaks
at $\lambda \leq 3.6$ cm and appears to be heavily absorbed at 13.3
cm, whereas flare V is effectively an optically thin event, brightest
at 13.3 cm and with no turnover evident in the radio spectrum. The
transition between these two extremes, through flares II -- IV, is
fairly smooth. Figures 1(b) \& (c) show how the spectral indices from
13.3 -- 3.6 cm \& 3.6 -- 2.0 cm respectively evolve with time :
$\alpha_{13.3-3.6cm}$ seems to evolve towards the mean optically thin
value (for Cyg X-3) of -0.55, whilst $\alpha_{3.6-2.0cm}$ seems to
reach even more negative values, suggestive of radiation losses
playing a role (see 3.3). Note that the spectral index from 13.3 - 3.6
cm was trivial to calculate as observations at these two frequencies
are made simultaneously; the spectral index from 3.6 - 2.0 cm was
calculated by averaging all 3.6 cm data points within 15 min of an
observation at 2.0 cm (similar methods with different binning periods
made little difference to the overall trends observed).

The transition towards lower opacity along the sequence I -- V is also
evident in the decreasing time delay between peak emission at 2.0 \&
3.6 cm and that at 13.3 cm. A delay in peak emission at wavelengths
longer than $\sim 6$ cm is often observed in Cyg X-3 flares and can
be explained as being due to emission at the longer wavelengths being
significantly absorbed at the time when the shorter wavelengths are
peaking. Only at later times when the ejecta have expanded
significantly for self-absorption (synchrotron or free-free, see e.g.
MPE92) to be unimportant, or for the ejecta to have travelled further
out in stellar wind (Fender et al. 1995) does the longer wavelength
emission peak. A decreasing time lag at 13.3 cm is thus a clear
indicator of decreasing opacity affecting the flares as the sequence I
-- V progresses.

\begin{table}
\caption{Time lag between peak emission at 3.6 \& 13.3 cm}
\label{symbols}
\begin{tabular}{@{}cc}
Flare & $\Delta$t \\
I & $\sim 24.5$ h \\
II & $\sim 12$ h \\
III & $ \geq 5$ h \\
IV & insufficient coverage \\
V & zero (no lag) \\

\end{tabular}
\medskip
\end{table}

Table 3 clearly illustrates the decreasing time delay of peak emission
at 13.3 cm relative to that at 3.6 cm in the sequence I -- V.  There
is a hint at times in the data set of a small time lag between
emission at 2.0 cm and that at 3.6 cm, but this effect is marginal and
no systematic trend is yet evident.

\subsection{The flare decay phase : evidence for radiation losses}

Hjellming, Brown \& Blankenship (1974) found that for the 1972
September radio flares the decline after peaking could be well
described during the initial few days by an exponential
decay. However, these authors found that after $\sim 4$ days the
decline in the radio was better fitted by a power-law decay of index
$\sim 4.9$.  The exponential decay proved troublesome for those
applying the simple van del Laan (1966) model for an adiabatically
expanding cloud of relativistic electrons, which predicts a power-law
decay at all times. However we note here that recent modelling work
(Canosa, Fender \& Pooley, 1997) has shown that the integrated flux
from an approaching and receding plasmon, each of which is decaying
with a power-law form in its rest frame, can mimic an exponential
decay.

MPE92 have tackled the reported change from exponential to power-law
decay in their model by invoking twin radio jets which expand in the
lateral direction exponentially at first, and then linearly at later
times. This reproduces well an initial exponential and subsequent
power-law decay.

In order to investigate the form of the decay of the radio flare
events the entire radio data set was searched by eye to look for
periods of relatively smooth flux decline.  The least-squares fitting
routine {\sl Gnufit} was then used to find a best fit to the decay
using both exponential and power-laws.  Radio observations of Cyg X-3
tend to show clearer variability at higher frequencies : this, coupled
with $\sim 3$ times more intensive coverage by the Ryle Telescope
during this period than at Green Bank, meant that there are
considerably more clear decays fitted at 2.0 cm than at 3.6 or 13.3
cm.  We fitted in total 9 clear decays at 13.3 cm, 12 at 3.6 cm and 27
at 2.0 cm.

In general the goodness of fit was very similar for both the
exponential and power-law decays, and it is difficult to confidently
discriminate between them. There is no evidence in the data for a
change from exponential to power-law decay of index $\sim 4.9$ after
$\sim 4$ days, as reported by Hjellming et al. (1974) (though this
would be hard to observe as the mean interflare interval was 4 - 5
days). We shall take the measured exponential decay constant, $\tau$,
below as being representative of the rate of decay at each wavelength.

The mean values for $\tau$ are $0.18 \pm 0.09$, $0.55 \pm 0.37$ \&
$1.01 \pm 0.26$ d at 2.0, 3.6 \& 13.3 cm respectively. Although the
uncertainties in the fits to the Green Bank data are large
(particularly at 3.6 cm), there remains strong evidence for
consistently longer decay times at longer wavelengths. This is
contrary to the simple model of van der Laan 1966, in which the sole
energy loss mechanism is adiabatic expansion and the same decay rate
is expected at all wavelengths.

\subsection{Enhanced and variable (sub)mm and infrared emission}

The JCMT (sub)mm observations of Cyg X-3 during 1994 Feb - March
reveal the source to be in a more active state than mm observations
during a period of radio quiescence in 1993 (Fender et al. 1995). There
appears to be a fresh injection of particles between the two
observations on JD 2448410, with a $77 \pm 20$ mJy detection at 0.8 mm
coming less than 20 min after a $3\sigma$ upper limit of 40 mJy at 1.1
mm was established. A significant change in flux density at $\sim 1$
mm on a timescale of $\sim 20$ min implies a limiting size to the
emitting region of $\leq 4 \times 10^{13}$ cm. This probably places
the emission near the base of the jet, well within the cm
`photospheres' (Fender et al. 1995, Waltman et al. 1997).  The detection
of Cyg X-3 at at flux density of $\sim 330 \pm 69$ mJy at 0.45 mm on
JD 2449412 shows the source to be anomalously strong at this
wavelength, where an extrapolation of the spectrum from 2.0 cm with
spectral index -0.55 would have predicted a flux density of $\sim 50$
mJy. This high flux at 0.45 mm may have been a precursor to flare II,
which peaked less than 24 hr later.

Cyg X-3 was also anomalously bright in the infrared H \& K bands, by
at least a factor of two, during the radio flaring period. On JD 2440406
the flux density was seen to increase rapidly during the observations.
This effect was strongest in the K band, resulting in a reddening of the
source in (H-K). This is similar to the effect seen during rapid infrared
flaring (e.g. Fender et al.. 1996), but given the rapid evolution of such
events, the non-simultaneity of the H- and K-band measurements does not
allow much information to be derived from this observations. Also, the
observations on JD 2449407 reveal the K-band flux density to be
decreasing steadily on a timescale of $\sim 10$ min. Combined with
comparable flux densities observed on the previous night, this implies
that some ongoing replenishment of the infrared-emitting material must
be taking place.

\section{Discussion}

\subsection{Radiation losses}

In the model of MPE92, wavelength-dependent
synchrotron and inverse Compton `radiation
losses' were not considered to be significant when compared to the
adiabatic expansion losses suffered by the laterally expanding
jet. However, there is strong evidence presented here that radiation
losses are important. This evidence is in the form of repeated
examples in the data set of more rapid flux density decays at shorter
wavelengths (section 3.3). It is particularly important that two
wavelengths on the optically thin branch of the synchrotron emission
(i.e. 3.6 cm \& 2.0 cm) consistently show different decay rates, as
decays at 13.3 cm can be clouded by evolution of the opacity of the
ejecta itself (MPE92 find a turnover wavelength of $\sim 6$ cm between
the optically thin and initially self-absorbing branches of the
emitted spectrum). The same interpretation can be drawn from observing
that $\alpha_{3.6-2.0cm}$ decreases below the mean optically thin
branch value of $\sim -0.55$ (Fig 1(c)).

At this point we will summarize the absorption and loss mechanisms
which may affect the observed emission from a radio-emitting plasmon
in an astrophysical environment.

\subsubsection{Absorption mechanisms}

\begin{itemize}
\item{Synchrotron self absorption.}
\item{Free-free absorption from thermal electrons either entrained 
in jet/plasmon or in surrounding environment (e.g. wind).}
\end{itemize}

In the model of MPE92, Synchrotron self-absorption is not significant
at {\em any} time for $\lambda \leq 6$ cm. Free-free absorption by
entrained thermal electrons could be occurring. Free-free absorption
by thermal electrons in a wind, as envisaged by Fender et al.. (1995)
to explain quiescent cm-mm emission, should not affect the larger
flares studied here as by the time decay commences ($\geq 1$ day) the
ejecta should have passed beyond the relevant radio `photospheres'
(and Waltman et al.. 1997 have in addition shown that the photospheric
radii used in Fender et al.. 1995 may have been overestimates).
Furthermore, free-free effects would attenuate {\em
long-wavelength} emission most strongly, producing the opposite effect to
that observed here; for this reason we do not consider these processes
in the following analysis (although they are undoubtedly important in
detailed modelling of flares).

\subsubsection{Electron loss mechanisms}

\begin{itemize}
\item{Adiabatic expansion losses.}
\item{Synchrotron losses.}
\item{Inverse Compton losses.}
\item{Bremsstrahlung between relativistic electrons and `thermal' 
protons.}
\item{Ionisation losses.}
\end{itemize}

Adiabatic expansion losses produce wavelength-independent decays and
were the only loss mechanism considered by van der Laan (1966). It
seems beyond doubt that some plasmon expansion is occuring, so this
process cannot be ignored (and is in fact the benchmark against which
the significance of other processes can be tested). Both Synchrotron
and Inverse Compton losses produce wavelength-dependent decays, with
shorter-wavelength emission being attenuated most rapidly as a result
of the higher energy electrons losing energy fastest. The crucial
parameters for these two wavelength-dependent mechanisms are magnetic
field and radiation field respectively. Together, we refer to these
processes as `radiation losses', and as their signatures are clearly
observed, we shall consider them in detail. The exact contribution
from Bremsstrahlung between the nonthermal electrons and thermal
protons is unclear, but as (a) we are not attempting to model the
thermal electron population and (b) these losses are wavelength
independent anyhow, we will not consider this process. Finally,
Ionisation losses are mildly wavelength-dependent but much less so
than either of the `radiation losses' and anyhow we expect the
environment of the plasmon to be effectively fully ionised, so again
do not consider this process.

Therefore, while it is an oversimplification, we conclude that the
most important processes to consider are adiabatic expansion, 
synchrotron and inverse Compton losses. The evidence has already
been presented for the significance of one of the latter two processes;
it remains to be seen which process is dominant, and over what
timescale.

\subsubsection{Modelling the electron loss mechanisms}

Following the work of MPE92, we retain their jet geometry and
formulation governing the evolution of the lateral expansion velocity,
jet radius and hence magnetic field (via conservation of magnetic
flux) with time.  From this the relative contributions from adiabatic
expansion losses and synchrotron losses can be calculated. We futher
make an approximation to the radiation field and calculate the
relative contribution of inverse Compton losses. Details of the
equations governing the model are given in Appendix A of Fender
(1995).  The model parameters to be fit are :

\begin{itemize}

\item{$B_0$, the magnetic field at the base of the jet}

\item{$t_c$, the time at which the lateral expansion changes from exponential
to linear}

\item{$t_e$, the time constant for the exponential expansion, related to
the observed decay constant $\tau$ by $t_e = \tau(7p-1)/6$ (where p is the
electron energy index, 2.1 for an optically thin spectal index of -0.55)}

\item{$T_*$ \& $R_*$, the temperature and radius of the black-body being used
to approximate the radiation field from the companion to the compact object.}

\end{itemize}

\begin{figure}
\centering
\leavevmode\epsfig{file=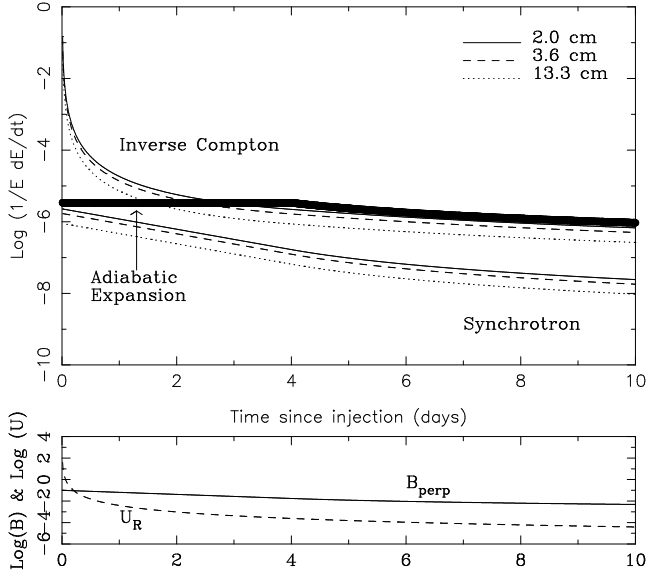, width=8.5cm, clip}
\caption{A comparison of loss mechanisms in the Cyg X-3 jet. Adopted
jet geometry follows MPE92; magnetic field at injection $B_0 = 0.1$ G;
luminous companion star (10 R$_{\odot}$, 40 000 K). Top panel plots
$E^{-1}$ dE/dT 
(which is a measure of the number of electrons of energy lost)
as a function of time as a result of adiabatic expansion, inverse Compton,
and synchrotron losses. The lower panel plots the evolution of 
$B_{\rm perp}$
(component of magnetic field perpendicular to jet axis) and $U_R$ (photon
energy density local to ejecta) with time. The transition from exponential
to linear lateral expansion is modelled, following MPE92,
to occur 4 days after injection.}
\label{}
\end{figure}

\begin{figure}
\centering
\leavevmode\epsfig{file=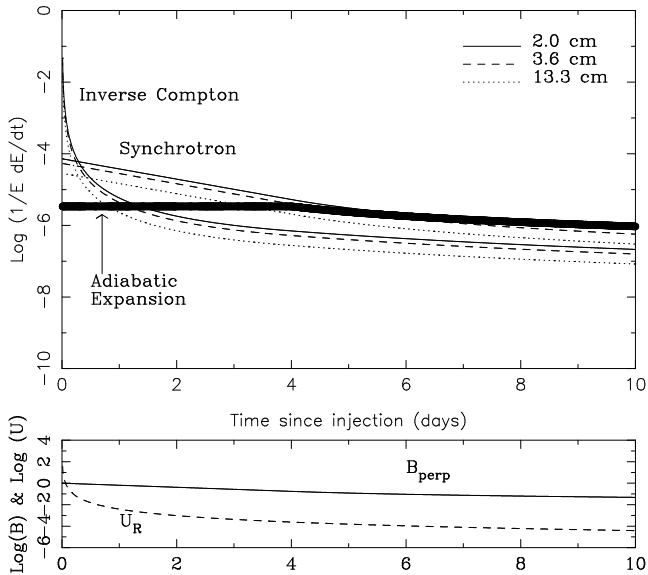, width=8.5cm, clip}
\caption{As for fig 3, but with $B_0 = 1.0$ G.}
\label{}
\end{figure}

\begin{figure}
\centering
\leavevmode\epsfig{file=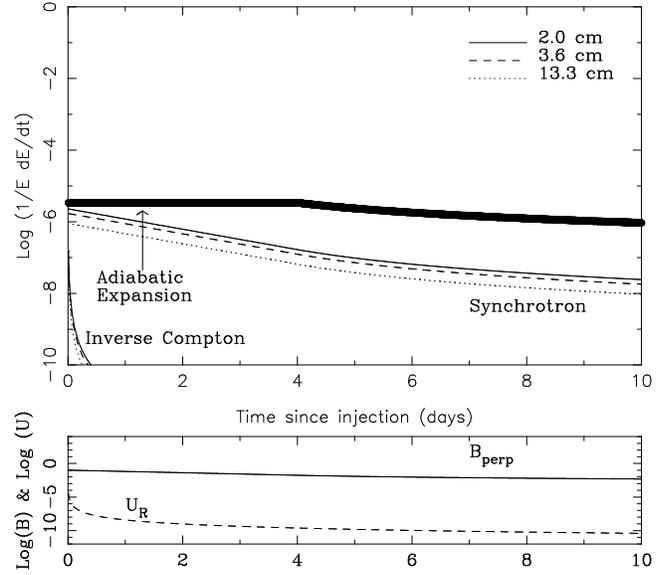, width=8.5cm, clip}
\caption{As for fig 3, but for main sequence (1 R$_{\odot}$, 4000 K) 
companion, i.e. LMXRB scenario}
\label{}
\end{figure}

From the observed change from exponential to power-law decay in the
1972 flares, we take $t_c$ to be 4 days, as did MPE92 (although there
is no evidence for this in the data, it may still have occurred, at
least for flares I -- IV, having been lost in the rise of the
following flare).  Assuming radiation losses to be of least
significance at 13.3 cm, we take the observed decay mean constant
$\tau = 1.01$ days at this wavelength and use it to derive a value for
$t_e$ of 2.31 d. We note here that while MPE92 predict some
self-absorption at 13.3 cm, the decay rate at this wavelength is
fairly consistent in all measurements (up to four days after flare
peak in some cases) whereas self-absorption is expected to be of
little significance after $\sim 2$ days. Infrared observations
(e.g. van Kerkwijk et al. 1996) indicate that the companion to the
compact object has an absolute magnitude in the near IR of $\leq -5$.
An approximation to such a luminous object can be made by taking a
black-body of $T_*$ = 40 000 K and radius $R_* = 10 R_{\odot}$. Note
that although the radiation field from the companion star almost
certainly comes from a far more extended region (i.e. a bright stellar
wind, particularly in the case of the companion being a Wolf-Rayet
star), given the high ejection velocity of $\sim 0.3$ c, the
approximation to a relatively small black-body should be adequate.

We use $B_0 \sim 0.1$G as fit to the data by MPE92 and also obtained
from applying equipartition theory to the Cyg X-3 radio jets by
Spencer et al. (1986) (we note however that there are no {\em a priori}
reasons for energy equipartition to be a valid assumption - see
e.g. Leahy 1991).

Using the above parameters, we find that inverse Compton losses
dominate the loss mechanisms for the first two days of jet expansion,
and after this time remain important at between 10 -- 50 \% of the
level of the expansion losses, which dominate from this point
onwards. The exact proportion of the contribution from radiation
losses is wavelength dependent, being most important at shorter
wavelengths.  Fig 3 plots $E^{-1} (dE/dt)$ (the inverse of the
electron lifetime - this is proportional to the number of electons of
energy E lost and is therefore a measure of the relative importance of
each loss mechanism) against time for this model.

Synchrotron losses do not play a major role, although increasing $B_0$
to 1.0 G is sufficient to make them important (Fig 4). This is not an
unphysical field and synchrotron losses should be considered in future
models of radio emission from Cyg X-3.

Several authors dispute the high-mass, luminous nature of the
companion star to Cyg X-3, preferring a low-mass X-ray binary (or
similar) scenario. The relative contributions of the three loss
mechanisms in such a scenario (where we represent the companion star
in this case by $R_* =$ R$_{\odot}$, $T_* = 4000$ K) are illustrated
in Fig 5.  Given that radiation losses {\em are} observed, then if the
LMXRB scenario is correct B$_0 \geq 1.0$ G.

We have found here that in adopting the geometry of MPE92 whilst
including a massive luminous companion star, inverse Compton losses
dominate for $\sim$ the first two days after injection. This is
inconsistent with the model fits put forward by MPE92 and in turn
casts some doubt upon their fits and geometry. However, it should be
noted here that inverse Compton losses are the most
geometry-independent of the loss mechanisms considered here and are
not strongly affected by small changes in assumed jet
configuration. In particular, a model was run for constant lateral
expansion immediately after injection (i.e. no period of exponential
lateral expansion) -- again inverse Compton losses dominated over
adiabatic and synchrotron losses when a highly luminous companion was
considered.  We should note here that the MPE92 model fits were for
the more luminous 1972 radio flare which may have had different
outflow/expansion parameters than the weaker events under discussion
here.

\subsubsection{$\gamma$-ray emission}

In the inverse Compton process, electrons of energy $\gamma m_ec^2$
will scatter photons of frequency $\nu$ to a mean frequency of $\sim
\frac43 \gamma^2 \nu$ (e.g. Hughes \& Miller 1991). Electrons
radiating synchrotron emission at 2.0 cm at the base of the jet (i.e.
where we assume B=0.1 G) will have energies of $\sim 0.1$ GeV and
hence Lorentz factors $\gamma \sim 200$. Thus inverse Compton losses
suffered by these electrons will cause photons in frequency range
$10^{14} - 10^{16}$ Hz (i.e. infrared through UV, where emission from
the luminous companion should be concentrated) to be re-emitted after
interaction with the electrons in the frequency range $5 \times
10^{18} - 5 \times 10^{20}$ Hz (i.e. 20 keV - 2 MeV) : hard X-rays
through $\gamma$-rays. Based upon this, and the observed importance of
radiation losses, probably inverse Compton, at 2.0 cm, we predict hard
X-ray/$\gamma$-ray bursts associated with radio flares. There may also
be some constant contribution to the hard X-ray/$\gamma$-ray flux from
the inverse Compton mechanism if the jet is continuous, but with a
lower injection rate, during quiescence.

\subsection{Decreasing local opacity}

\begin{figure*}
\begin{minipage}{177mm}
\centering
\leavevmode\epsfig{file=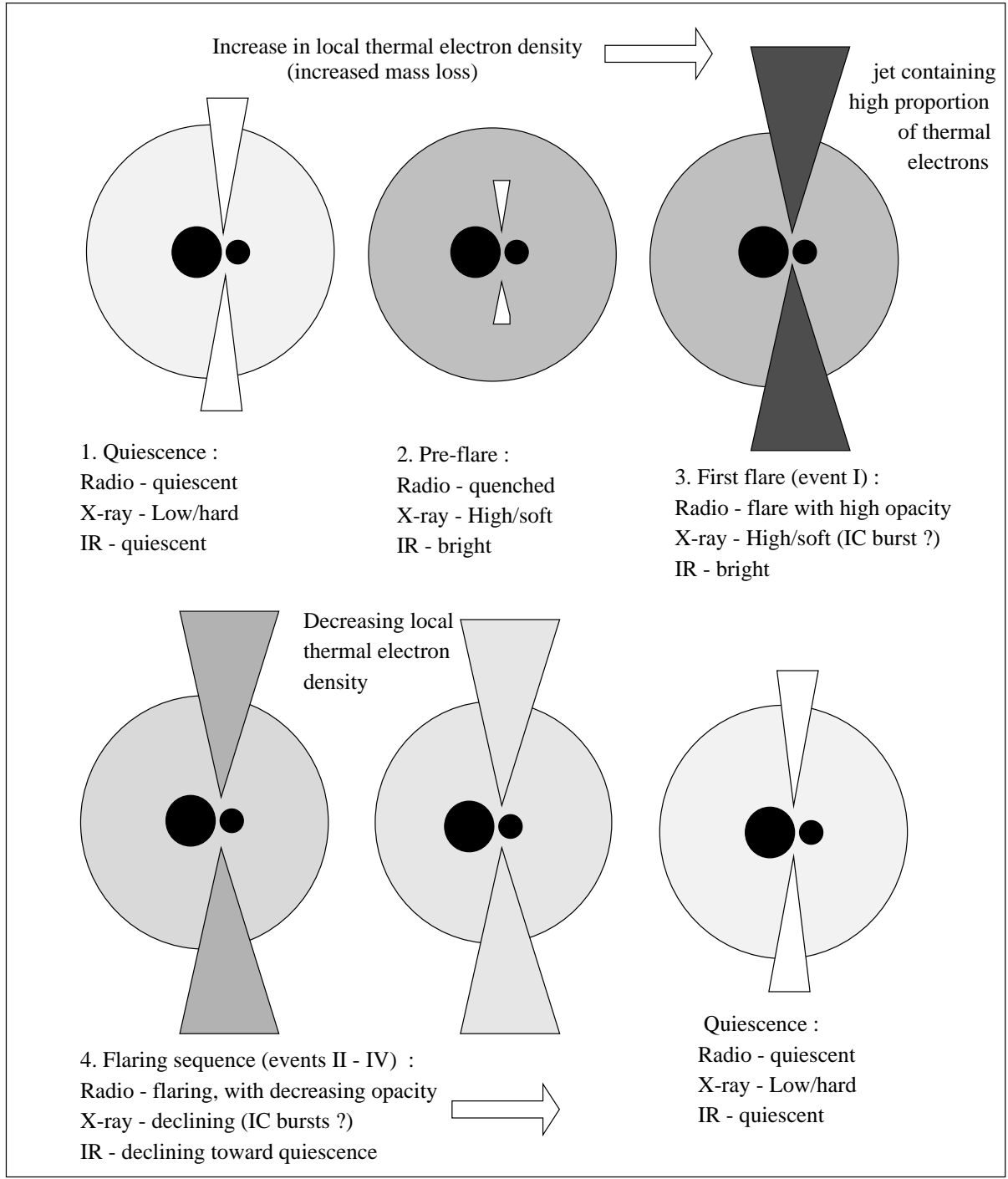, width=16cm, clip}
\caption{Decreasing local opacity during four or five large jet events.
Increased mass loss by companion star causes quenched radio emission but
bright infrared and X-ray fluxes. Eventually a large injection of 
relativistic particles into the jet occurs, entrained with which are a
high proportion of absorbing thermal electrons. As the mass loss rate
declines toward its quiescent level, a series of radio flares with 
progressively lower opacity occur. During this period the infrared and
X-ray brightnesses decline gradually toward their quiescent levels. Radio
flares may also be accompanied by X-ray/$\gamma$-ray bursts due to inverse
Compton interactions between optical photons from the companion star
and high energy electrons in the jet.}
\label{}
\end{minipage}
\end{figure*}

Given the clear trends, in particular the decreasing opacity, observed
during the flaring sequence, we have attempted to construct a testable
scenario for the evolution of the outburst. Our favoured scenario is
based upon a mass transfer (mass loss) instability, although other
scenarios are considered.

\subsubsection{Mass transfer (mass loss) instability scenario}

In this scenario we consider that events I - V (or at least I - IV) are
as a result of separate injection events, and that the
evolving characteristics reflect changing condition in the region of
injection.  The scenario is envisaged as follows, and illustrated 
in Fig 6.

\begin{itemize}
\item{Stage 1:
Initially Cyg X-3 is in a quiescent state, with a low-level jet 
responsible for the quiescent radio emission.}
\item{Stage 2:
A period of increased mass-loss
from the companion star results in an increasing density of thermal electrons
in the region of the accreting compact object. 
This causes the quenching of
the jet (exact mechanism unclear)
and thus radio emission is reduced. At the same time the infrared
flux (proportional to $n_e n_i$ where $n_e$ \& $n_i$ are the electron and
ion number densities) increases due to the higher electron and ion number
densities, as does the X-ray flux, due to a higher accretion rate.}
\item{Stage 3:
Radiation and/or magnetic pressure builds up in the inner regions of the
disc (the exact mechanisms for producing jets remain unclear) and finally
there is an explosive injection of relativistic particles into the jet.
This produces a radio flare, with considerable associated opacity due to
a high proportion of absorbing thermal particles entrained in the jet.}
\item{Stage 4:
Subsequently the mass-loss rate from the companion star subsides towards
its normal levels and a sucession of injections with decreasing proportions
of associated absorbing thermal electrons occur. 
During this period the infrared and
X-ray brightness begin to decline toward their normal levels (although
there may still be X-ray/$\gamma$-ray flares due to inverse Compton losses
from later flares). Eventually the source returns to quiescence c.f.
stage 1.}

\end{itemize}

Flare V is anomalous and may 
be due to an interaction between the ejecta and part of the ISM. If not,
then it would appear that the rate of injection of relativistic particles
into the jet had declined dramatically by the time this event occurred,
resulting in a much longer rise time.

This scenario has a lot of attractions in that it seems to explain the
observed evolution of the opacity in the sequence I -- V, {\bf plus} the
strong infrared brightness and observed correlation between radio flares
and bright X-ray states (Watanabe et al. 1994). It is also
consistent with the suggestion from the infrared and (sub)mm data that
continued injections were taking place {\sl at least} between flares I \& II.

\subsubsection{Alternative scenarios}

In attempting to explain the decreasing opacity during the flare
sequence several other possible scenarios were considered, in
particular one in which the flare events were recombination shocks
occuring at increasingly large distances along the jet, and hence away
from a source of opacity (the stellar wind from the companion). There
is evidence for such shocks causing {\em in situ} acceleration of
particles in the jets of AGN.  However the assumed wind geometry
(c.f. Fender et al. 1995) and jet velocity (0.3 c, c.f. Schalinski et
al 1995) are inconsistent with the rather slow evolution of the
opacity in this scenario.  Furthermore the multiple-shock scenario is
clearly inconsistent with our conclusion that inverse Compton losses
are the dominant radiation loss mechanism, and would require the
ejecta to maintain a high magnetic field ($\sim 1$ G) and lose energy
via synchrotron losses.

If Cyg X-3 turns out to have a low-mass companion, as proposed by
several authors (e.g. White \& Holt 1982; Mitra 1996) then it seems
unlikely that a mass-transfer rate instability could cause the
oubursts. Such instabilities are most likely in systems in which
accretion via a stellar wind is dominant. In a low-mass, Roche-lobe
overflowing scenario then it may be that disc instabilities, such as
those proposed to account for X-ray transient events, are responsible
for the outbursts.

\section{Conclusions}

We present detailed observations of an outburst from the
enigmatic and poorly-understood X-ray binary Cyg X-3, revealing
new phenomena and helping to clarify the intrepretation of 
previous observations across a broad wavelength range. 

We report for the first time confirmation of strongly quenched radio
emission prior to outburst, independent of observations at Green Bank
(Waltman et al. 1994, 1995). The quenched period, lasting $\sim 19$
days in the outburst reported here, is clearly an important period
when the physical conditions local to the base of the jet/accretion
disc region have changed significantly.  Whether the quenching is due
to changes in the accretion disc or to physical quenching by
circumstellar material is unclear, and further study of this phase is
undoubtedly important.

Detailed observation and measurement of characteristic decay times at
three wavelengths reveal conclusive evidence for wavelength
dependence. Measurements at 3.6 \& 2.0 cm are particularly significant
as these lie on the optically thin tail of the synchrotron emission
and self-absorption does not have to be taken into account (as it does
in most cases at 13.3 cm). We fail to confirm the transition from
exponential to power-law decay after $\sim 4$ days reported by
Hjellming et al. (1974), although this may have occurred and been lost
in the rise of a subsequent flare. The clearest interpretation of
wavelength-dependent characteristic decay times is that radiation
losses are important processes in the Cyg X-3 jet.  The same
conclusion was drawn by Baars et al. (1986) and Fender et al. (1995)
based upon comparison of mm and cm observations, and is consistent
with the rapid fluctuations we have observed at (sub)mm wavelengths
during the outburst.  In order to investigate this phenonemon we have
calculated (wavelength-dependent) synchrotron and inverse Compton
losses as well as (wavelength-independent) adiabatic expansion losses
using the geometry of MPE92. We find that in the presence of a strong
radiation field such as that from a luminous companion star, inverse
Compton processes are the dominant loss mechanism for the first $\sim
2$ days after injection of particles into the jet. Without a luminous
companion, as in a LMXRB scenario, a magnetic field of at least 1.0 G
is required at the base of the jet.

The opacity in the radio emitting region (assumed to be the jet), as
measured both by spectral indexes and time lags between emission at
3.6 \& 13.3 cm, is observed to be anomalously high at the beginning of
the flare sequence, declining gradually toward quiescence through the
sequence of flares I - V. Waltman et al. (1995) have found that the
highest opacity occurs immediately following quenching, consistent
with the very high opacity in flare I which occured less than 24 hr
after the last observed quenched fluxes. The apparent build-up of
opacity during quenching is suggestive of an increase in absorbing
thermal electrons in the vicinity of the radio emission, possibly as a
result of an enhanced density of the stellar wind from the companion
star.

Combining our investigations into the nature of Cyg X-3 outbursts with
ideas presented in Kitamoto et al. (1994), Watanabe et al. (1994) and
Waltman et al. (1995), we have constructed a qualitative model for the
evolution of an outburst in Cyg X-3. We propose that the heightened
opacity in the radio emission is due to an enhanced proportion of
entrained absorbing thermal electrons, as a result of an increase in
density of the stellar wind from the companion star.  The build up of
opacity ceases at the last point of quenched emission and we observe a
sequence of radio flares as the system seeks to rid itself of the
build up of matter, possibly in the accretion disc.  During the entire
period the accretion disc has been bright and the source is in its
high and soft state (higher accretion rate plus more
scattering). Radio flares may be accompanied by hard
X-ray/$\gamma$-ray bursts as infrared and optical photons are
upscattered to higher energies via inverse Compton processes.  The
bright but declining infrared state of Cyg X-3 as observed during the
outburst is in agreement with this model, though we cannot
discriminate on the basis of our observations between possible
contributions from a bright accretion disc and an enhanced stellar
wind. Further multiwavelength observations of
Cyg X-3 in outburst are clearly required to test and refine this
scenario.

\section*{Acknowledgements}

We would like to thank Josep Mart\'{\i} for many useful comments and
discussions relating to this work, the observers at UKIRT and JCMT for
giving up time to make observations at short notice, and Mark Garlick
for proof-reading an early draft. RPF would like to thank the Open
University for a research studentship during the period of this
research.  Radio Astronomy at the Naval Research Laboratory is
supported by the Office of Naval Research.  The Ryle Telescope is
funded by the UK Particle Physics and Astronomy Research Council
(PPARC).  The UKIRT is operated by The Observatories on behalf of
PPARC. The JCMT is operated by The Observatories on behalf of PPARC,
The Netherlands Organisation for Scientific Research, and The National
Research Council of Canada.  This research made use of the SIMBAD
database, operated at CDS, Strasbourg, France.


\end{document}